\documentstyle[12pt]{article}
\input{psfig}
\begin{document}
\baselineskip 23pt
\bibliographystyle{unsrt}
\pagestyle{plain}
\title{Self-Wiring of Neural Networks}
\author{Ronen Segev and ${}^1$Eshel Ben-Jacob\\
School of physics and astronomy,\\
 Raymond \& Beverly Sackler Faculty of exact Sciences,\\
Tel-Aviv University, Tel-Aviv 69978, Israel \\
\(1\) communicating author,\\
email: eshel@venus.tau.ac.il,\\
telephone no.:972-3-6407845,\\
Fax: 972-3-6422979}

\maketitle

\begin{abstract}
In order to form the intricate network of synaptic connections in the brain, the growth
cones migrate through the embryonic
environment to their targets using chemical communication.
As a first step to study self-wiring, 2D model systems of neurons have been used. 
We present a simple model to reproduce the salient
features of the 2D systems. 
The model incorporates random walkers representing the growth cones, which
migrate in response to chemotaxis substances extracted by the soma and
communicate with each other and
with the soma by means of attractive chemotactic "feedback". 
\end{abstract}         
Pacs no.: 87.22.-q  05.70.Ln  82.20.Mj.

The intricate network of connections between neurons has a crucial effect on the
information processing ability of nervous systems \cite{TL96,TN91,MI89}.
The precise pattern of the synaptic connections is essential for proper functioning of the system.
The task of self-wiring during embryo genesis is perhaps one of the most
staggering examples of self organization in complex systems.
In a human brain, for instance, there are approximately
\( 10^{11} \) neurons that form \( 10^{16} \) synaptic connections.

Neuronal connections are formed when the growth cone of each neurite migrate from their
site of
origin, on the neuronal soma, through the embryonic environment to their synaptic target. 
The growth cones navigate using sophisticated means of chemical signaling for communication
and regulations, and by molecular guidance cues introduced into the environment by the
different cells, (e.g. neurons soma, glia cell: astrocytes, and oligodendrocy)
\cite{TL96,TL91,DJ88,BU94,GB79}.
The wiring process proceeds by emission of neurites (at this stage we refer to either
dendrites or axons as neurites) from the neuronal soma. Each neurite is tipped
by a growth cone. The growth cone is capable of measuring concentration and
concentration gradients of substances in the environment. 
It is composed of a central core which is an extension of
the neurite itself, and is rich in microtubulets that provide the structural
support. The core is rich in mitochondria, endoplasmic reticulum and
vesicular structures. Surrounding the central core are regions known as
lamellipodia, in which the contractile protein Actin is abundant. At the extremities
of the lamellipodia there are very thin straight filaments known as filopodia. The
filopodia are in constant motion, as they extend from the lamellipodia and retract back
to it.
The growth of the neurite occurs when filopodia extend from the lamellipodia and
remains extended rather than  retracts as the end of the lamellipodia advances
towards the filopodia.
The complexity and the dynamics of the growth cones hint that they might act as 
autonoumus entities. Indeed, there are direct experimental observations of the activity of
growth cones separated from their neurites, that support this view \cite{HM84}. The above  
observations are essential to the construction of our model.
In particular, they led us to represent each of the growth cones as an entity (walker)
with its own internal energy as described below.

Clearly, self wiring of the brain is far too complicated to be the first  problem to study. Hence much effort is
devoted to in vitro experiments of simpler  2D model systems \cite{TL96}.  In these experiments neurons are
placed on a PLL surfaces so it is easier to monitor their self-wiring \cite{BR97,Dwir96}. Here we propose a model to
describe self-wiring in such a systems.

Microscope observation reveal that \cite{BR97} the movement of the growth
cones appears to be a non-uniform random walk with the highest probability to move forward ("inertia")
and high probability to move backward ("retraction"). The movement is not continuous, as
there are time intervals during which the growth cones do not move. The growth rate is of
the order of micrometer per minute \cite{GB79}. Extensive studies in vivo and in vitro 
revealed that the movement of neurites can be affected by four types of guidance cues: attractive or repulsive cues 
that can be either short-range or long-range. The short-range cues are contact mediated by non diffusive cell 
adhesion molecules (CAM) and extra cellular matrix (ECM) molecules. The long-range forces are mediated by 
emission of chemoattractant and chemorepellent substances which "pull" and "push" the growth cone
from the soma or its neurites \cite{TL96}. Clearly, the repulsive and attractive mechanisms should not
affect the movement of the growth cone simultaneously. Our assumption is that the
relative sensitivity (magnitude of response) to the two mechanisms is determined by the metabolic state of the
growth cone,  as we discussed below.

Our model of self wiring is inspired by the communicating
walkers model used in the study of complex patterning of bacterial
colonies \cite{EB94,EB95}. We assume the existence of two chemotaxis agents, 
repulsive-R and attractive-A, and a triggering agent T.
While the existence of chemotactic response has been demonstrated \cite{TL96}, there are no direct observations of a
triggering agent. Our assumption about its existence is motivated by the use of triggering agents in other biological
systems \cite{EB95,KL93} and experimental observation that indicates 
the growth cones can change it`s response
to chemotactic substances during the growth process \cite{Song97}. 

The role of each agent field is described below.  
The concentration fields of the chemicals are modeled by solving the corresponding continuous
reaction diffusion equations on a triangular lattice of a lattice constant
\( a_0 \). We represent the neurons (each  composed of cell soma,
neurites and growth cones) by simple active elements that capture the
generic features of the neurons described above. Each of the soma is a
stationary unit occupying one lattice cell at \( \vec{R_j} \). It sends
neurites one at a time, and "feeds" them with internal energy, as specified below. The soma also continuously
emits chemorepellent and
emits a quanta of chemoattractant when it senses a triggering field
concentration above a threshold level. The neurites are simply defined as
the trajectories performed by the growth cones and are characterized by
their lengths \( L_i \).

Each growth cone is represented by an active walker specified by its
location \( \vec{r_i}\), it's internal energy \( E_i \) and its previous
step direction \( {{\theta}_i}^p \).
The assignment of internal energy to describe the metabolic state of the growth cone is the most crucial assumption in
our model. The assumption was first motivated by our modeling of bacterial colonies \cite{EB94,EB95,MA97} in which such  
internal energy turned out to be a crucial feature in modeling systems composed of biological elements. The assumption is supported
by two experimental observations concerning the The growth cones: 1. they are rich with mitochondria \cite{TN91}. 2. They can function after being
separated from their cell soma \cite{HM84}. Naturally, we assumed that the soma feed the
growth cone with internal energy, which the growth cone utilized for its metabolic processes. We further assumed that
the neurite consumes internal energy proportional to its length. The time evolution of the internal energy is given by :
\begin{equation} \frac{\mbox{d}E_i}{\mbox{d}t}=\Gamma_i (N_j)-\Omega-\lambda L_i +K(A)A
\end{equation}
Where \( \Gamma (N_j) \) is the rate of internal energy supplied by the
soma. It is a decreasing function of \( N_j \), the number of neurites sent out by the
soma. The growth cone consumes internal energy at a rate \( \Omega \), 
and its neurite consumes the internal energy at a rate \( \lambda \) per
unit length. The last term on the right hand side of eq. (1) describes the
absorption of chemoattractant by the growth cone. We assume (as is usually the case \cite{TL96}) that the
chemoattractant agent can be used by the growth cone as an energy source. \( K(A) \) is already measured in units of energy.

The soma supplies energy at a higher rate than the consumption rate \( \Omega \). 
Hence initially (short neurite's length \( L_i \)) the internal energy
increases. In the absence of chemoattractant, for 
\begin{equation}
\begin{array}{rl}
L_{i} > l_{c}  & \mbox{where } l_c\equiv (\Gamma - \Omega )/ \lambda  
\end{array}
\end{equation}
the internal energy decreases. When \(
\frac{\mbox{d}E_i}{\mbox{d}t} \) first becomes negative, the growth cone extracts a quanta of
triggering material. There is a refractory period \( \tau_T \) before
another quanta is extracted. If during \( \tau_T \) the growth cone
senses sufficient concentration of chemoattractant, or \( \frac{\mbox{d}E_i}{\mbox{d}t}
\) becomes positive, it will not extract another quanta of the triggering
agent. If \( \frac{\mbox{d}E_i}{\mbox{d}t} \) is negative for a sufficiently long time, so that
\( E_i \) drops to zero, the neurite and its growth cone degenerate and
 are removed. The growth cone responds to a triggering field (sent by another
growth cone) by emitting a chemoattractant, provided its internal energy
is above a minimum value. When the growth cone reaches another cell or
another cell's neurite, it creates a synaptic connection and its metabolic
processes are stopped.

Each walker performs off-lattice random walk of step size
$d$, at an angle \( \theta_i \) which is chosen out of 12 available directions.
Thus it moves from its location \( \vec{r_i} \) to a new location \( \vec{r'_i} \)
given by :
\begin{equation}
\vec{r'_i}=\vec{r_i}+d(cos \theta_i,sin\theta_i)
\end{equation}
At each time step (in the absence of chemotaxis), the walker first chooses one of
the
directions \( \Phi_i^{(n)} = (n-1)\pi /6 \), n=1,2, ... ,12, (\( \Phi_i^{(n)} \) is defined
relative
to the previous direction of movement \( \theta_{i}^{p} \)) from a non uniform
probability distribution \( P_o (n) \) shown in Fig (1b).
The higher probability is to continue to move in the same direction and to move backward.
The walker does not move every time step. After \( \Phi_{i}^{(n)} \) is selected,
a counter for that chosen direction (given n) is increased by one. 
The walker performs a movement only after one of the counters reaches a specified
threshold \( N_C \). The movement is in the direction \( \theta_i \) which corresponds
to that counter. This process acts as a noise reduction mechanism, in agreement with the
experimental observations.

In the presence of chemotactic materials, the probability distribution \( P_0 (n) \) (the
relative probability of choosing from the available 12 directions) is modified. The new
probability of moving in the n-th direction is given by :
\begin{equation}
P(n) = P_0 (n) + G_A \cdot S(A) \nabla_n A - G_R \cdot S(R) \nabla_n R
\end{equation}
Where \( A \) and \( R \) are the concentrations of chemoattractant and chemorepellent,  respectively.
\( \nabla_n \) is the directional derivative in the appropriate direction.
The functions \( S(A) \) and \( S(R) \) are prefactors which decrease for both high and low
concentrations.
\( G_R \) and \( G_A \), which determine the relative magnitude of response to 
chemoattractant and the chemorepellent, are functions of \( \frac{\mbox{d}E_i}{\mbox{d}t} \).
We assume that the growth cone is more sensitive to the chemorepellent while it
is close to its cell soma. Since \( \frac {\mbox{d}E_i}{\mbox{d}t} \) decreases with the neurites
length,
we simply assume here that \( G_R \) and \( G_A \) are decreasing and increasing functions of
\( \frac{\mbox{d}E_i}{\mbox{d}t} \) respectively. 
To complete the model we handle the corresponding continuous reaction-diffusion equations
for the chemical concentrations.
The equation for the chemorepellent concentration  R is given by :
\begin{equation} 
   \frac{ \partial R}{ \partial t}=D_R \nabla^2 R - \lambda_R R + \Gamma_{R} \sum_{soma}
\delta ( \vec{R} - \vec{R_j} )
\end{equation}
where \( D_R \) is the diffusion coefficient, \( \lambda_R \) is the rate of spontaneous decomposition of 
\( R \),  and \( \Gamma_R \) is the rate of
extraction of
R by the soma located at \( \vec{R_j} \). Similar equations are written for the
chemoattractant \( A \) and the triggering field \( T \) with the appropriate source terms according
to the
properties described above. We assume that \( D_R \) and \( D_A \) are of the same
order. We further assume that \( \lambda_{R} < \lambda_A \), so the
chemorepellent is long range with respect to the chemoattractant.  

As we have mentioned, the reaction-diffusion equations are solved on a triagonal lattice with a
lattice constant \( a_0 \). In the simulations \( a_0=1 \) and length is measured in units
of \( 10 \mu m \). This way, the fact that the soma occupies one lattice cell is in
agreement with the typical size of the neurons soma. In the experiments, the typical
distance between soma cells is up to \( 500 \mu m \) \cite{Dwir96,BR97}. In the simulations the typical
distance is about \( 20-50 \) \( a_0 \). Here it is feasible to simulate networks
composed of up to 500 neurons. A typical diffusion coefficient \( D \) of the chemicals is of the
order of \( 10^{-5} \) \( cm^2/sec \). Numerical stability of diffusion equations imposes
that the time step \( \Delta t \) will satisfy \( \Delta t < 0.25 a_{0}^{2}/D \). In the
simulations time is measured in units of 10sec and the dimensionless diffusion
coefficients are of the order of 10 so the step size is \( \Delta t=0.025 \).

Note that \( \Delta t \) is the basic time step for solving the reaction diffusion
equations. The basic time step of the growth cones (the choice of the probabilities \( \Phi_i (n) \) es.)
\( \Delta \tau \) is \(40 \Delta t \).
To test the consistency of the model and its
agreement with reality we compare the rate of advances of the growth cones with that of
the walkers. The measured rate is about \( 1 \mu m/minute \), which is in agreement with
the walkers rate of growth of one lattice constant in about \( 60 \Delta \tau \).

The structure of the non-uniform probability \( P_0 (n) \) (which includes the effects
of "inertia" and "retraction") is shown in Figure 1a. In Figure 1b we show typical
trajectories (neurites) of the growth cones emitted from the soma and migrating outward
under the influence of the chemorepellent extracted. To demonstrate the efficiency of
target-finding by the walkers, we show in Figure 2a simulations of a system composed of two cells. One is a "normal" cell which emits
neurites, while the other is a "mutant" which is incapable of emitting neurites but otherwise responds
normally to chemicals. We see that even neurites which originally have migrated
away from the target cell change their path and migrate towards this cell, once the target cell is triggered to emit a chemoattractant.
The self-wiring efficiency is demonstrated in Figure 2b, in which we simulate a system composed of two "normal" cells.

Recently, there have been experimental studies of the effect of imposed 2-fold anisotropy
on the wiring process \cite{Dwir96}. To mimic the imposed anisotropy we include in the model a comb of
lines, along which the growth cones have higher probability to move. The effect of such
imposed anisotropy, both parallel and perpendicular to the line connecting the two
cells, is shown in Figures 2c and 2d respectively.

Next we show how the soma cell can regulate
the wiring process. In Figure 3 we show a system of five cells, one normal at the center and four
"mutant" cells at the corners. All the parameters in Figures 3a and 3b are the same, but in
Figure 3b the soma cell at the center "feeds" the neurite at a higher rate.
As a result, the growth cones trigger target cells when they are further
away from their soma, and the soma is wired to all four neighbors and not only to two, as
is the case in Figure 3a.

In Figure 4 we show simulations of 13 cells systems. Again, only the
central cell is "normal" and all the others are "mutant" cells.
Initialy the central cell "feeds" the neurites at a rate such that the critical
length \( l_c \) eq.(2) corresponds to about half the distance between
the soma cells. As expected, in this case the soma is wired only to the
nearest neighbors.

After the central cell is wired to two neighbors, it doubles the "feeding" rate
of the neurites. Thus, \( l_c \) doubles and the neurites migrate past the nearest neighbors before they
first emit a triggering agent. As a result the central cell is wired to the NNN cells. We 
demonstrated how the generic features of chemotactic "feedback" and regulation via the rate of "internal"
energy "feeding" enable the neurons to perform the complex task of self wiring. Our model has two prediction to be tested in experiments:
1. The existence of triggering agent. 2. Since the growth cone can "feed" itself by the chemoattractant agent in the media, we expect that the growth cone
will repel from other cell's soma when the media is enriched artificially with chemoattractant.   
Here we studied nervous
systems composed of a small numbers of neurons. In larger systems we expect additional "feedback"
mechanisms to regulate the rate of "internal" energy "feeding". What we have in mind is
that the soma cells, combined with the extracellular matrix, act as an excitable media which supports
chemical waves and spiral waves \cite{SB98}. The rate of "feeding" is assumed to be tuned by these waves
and the electrical activity of the cells. 
\section*{Acknowledgments}
We have benefited from fruitful discussions with E. Braun, A Czirok, B. Dwir, D. Horn, D. Michaelson and D.
Kessler. We thank E. Braun for sharing with us his unpublished results. This research has been
supported in part by a Grant from the BSF no 95-00410, a Grant from the Israeli Science Foundation, and
the Sigel Prize for Research. 

\bibliography{physlettref}

\newpage

\begin{figure}[t]
\centerline{\psfig{figure=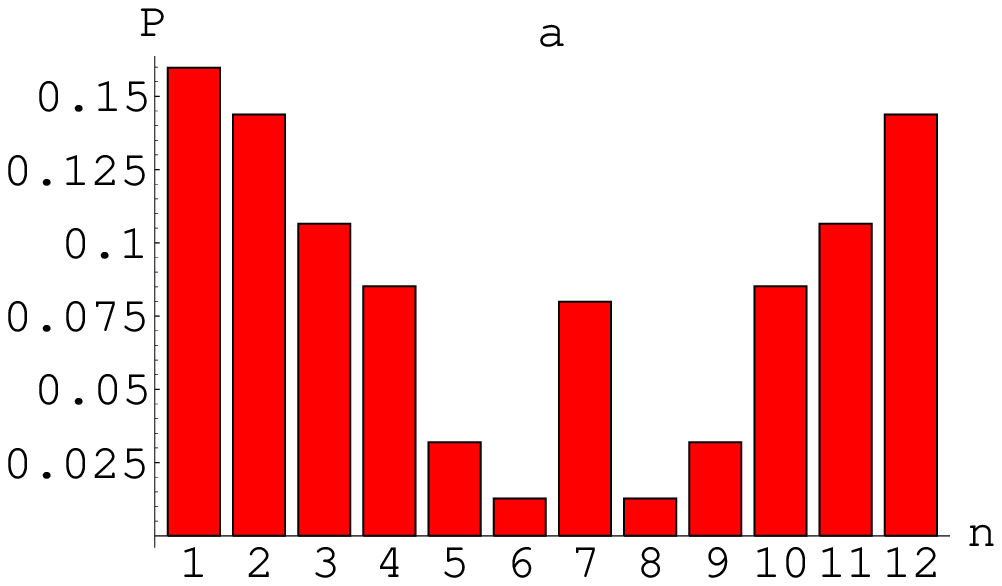,width=8cm,height=8cm,clip=} \psfig{figure=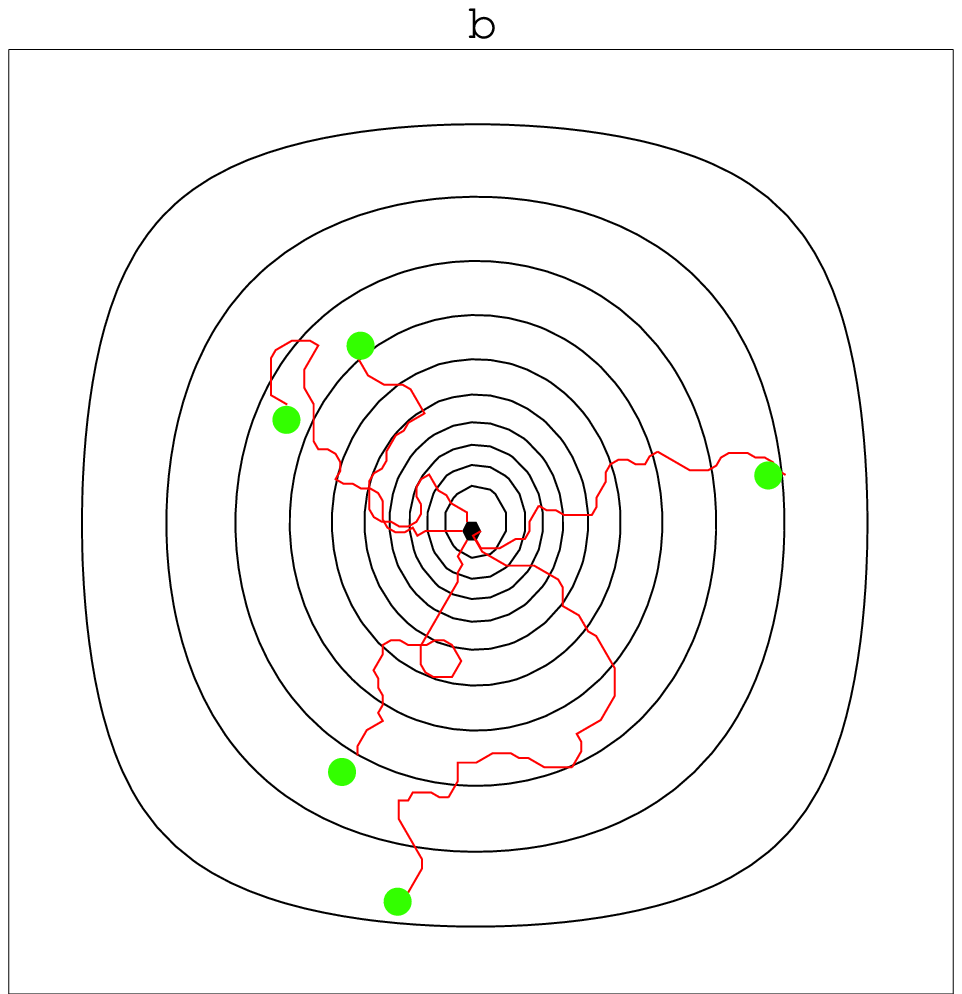,width=8cm,height=8cm,clip=}}
\caption []{
a. The non-uniform probability distribution \( P_0 (n) \). The highest probability is for
n=1, i.e to continue in the same direction ("inertia"). There is also high probability for n=7,
i.e to move backward ("retraction"). 
b. Migration of the walker under the influence of chemorepellent extracted from the soma
cell at the center. 
}
\end{figure}

\begin{figure}[btp]
\centerline{{\psfig{figure=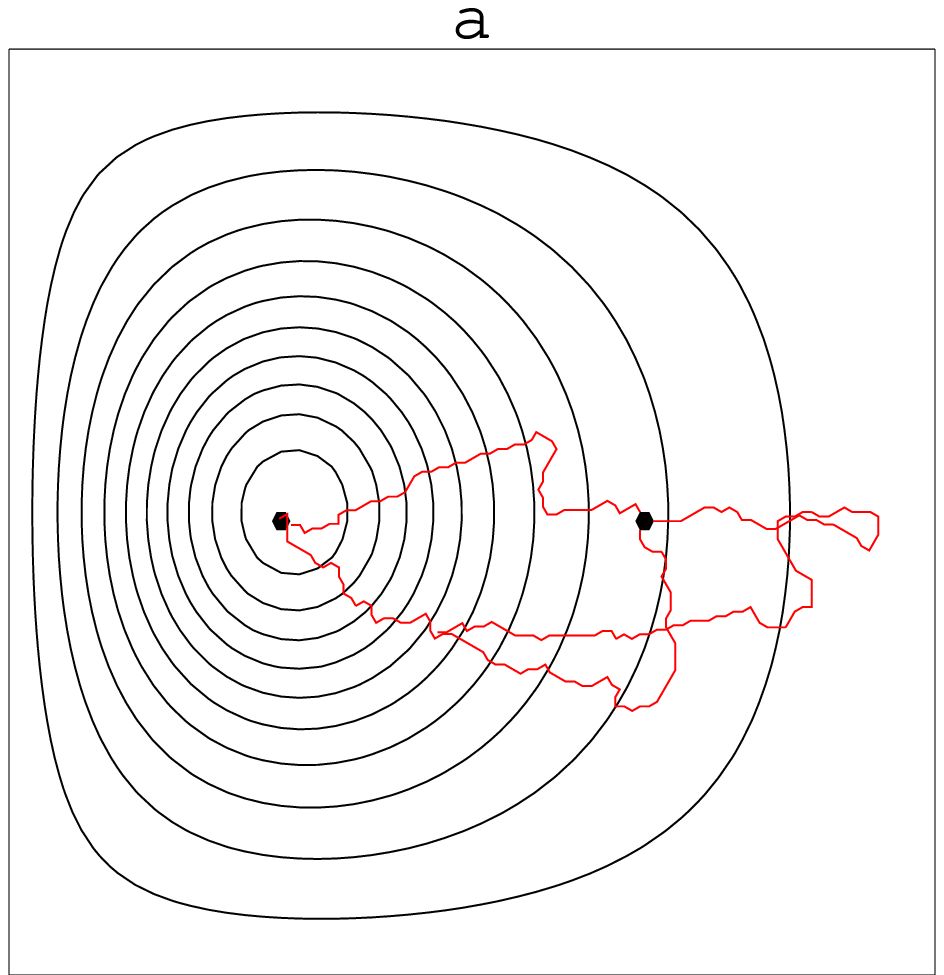,width=8cm,height=8cm,clip=} \psfig{figure=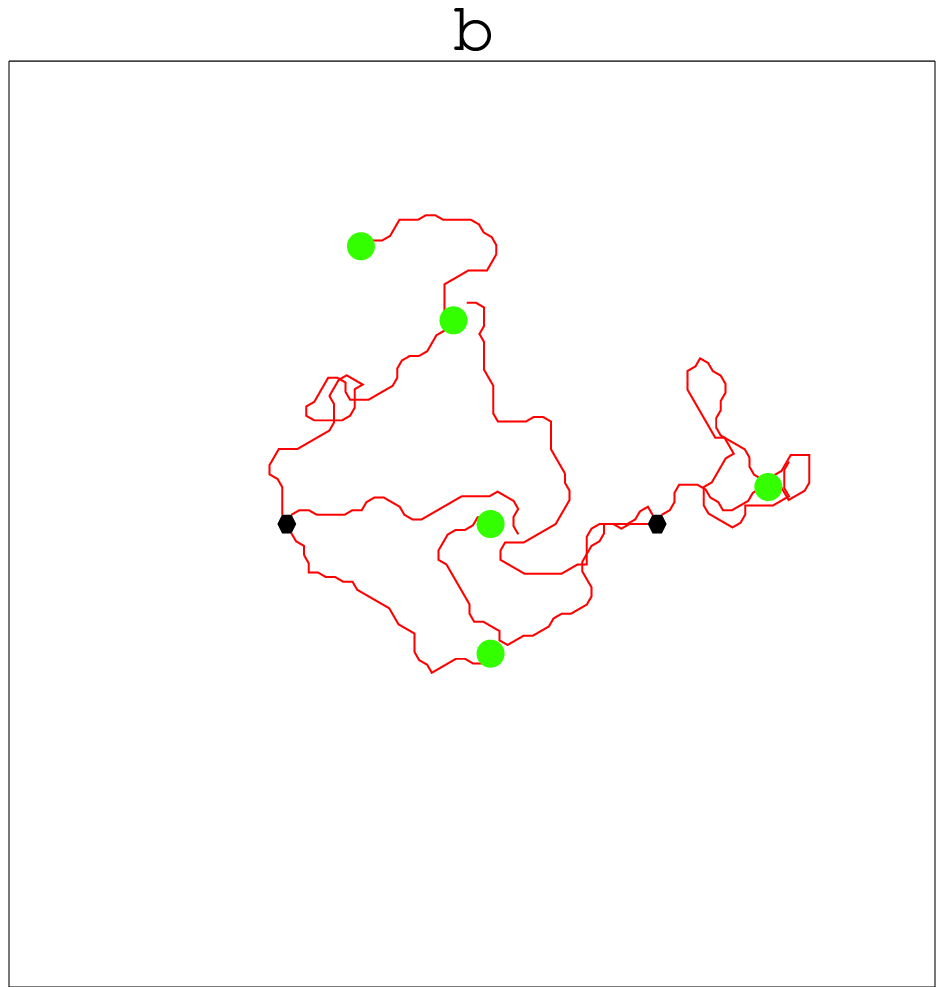,width=8cm,height=8cm,clip=}}}
\centerline{\psfig{figure=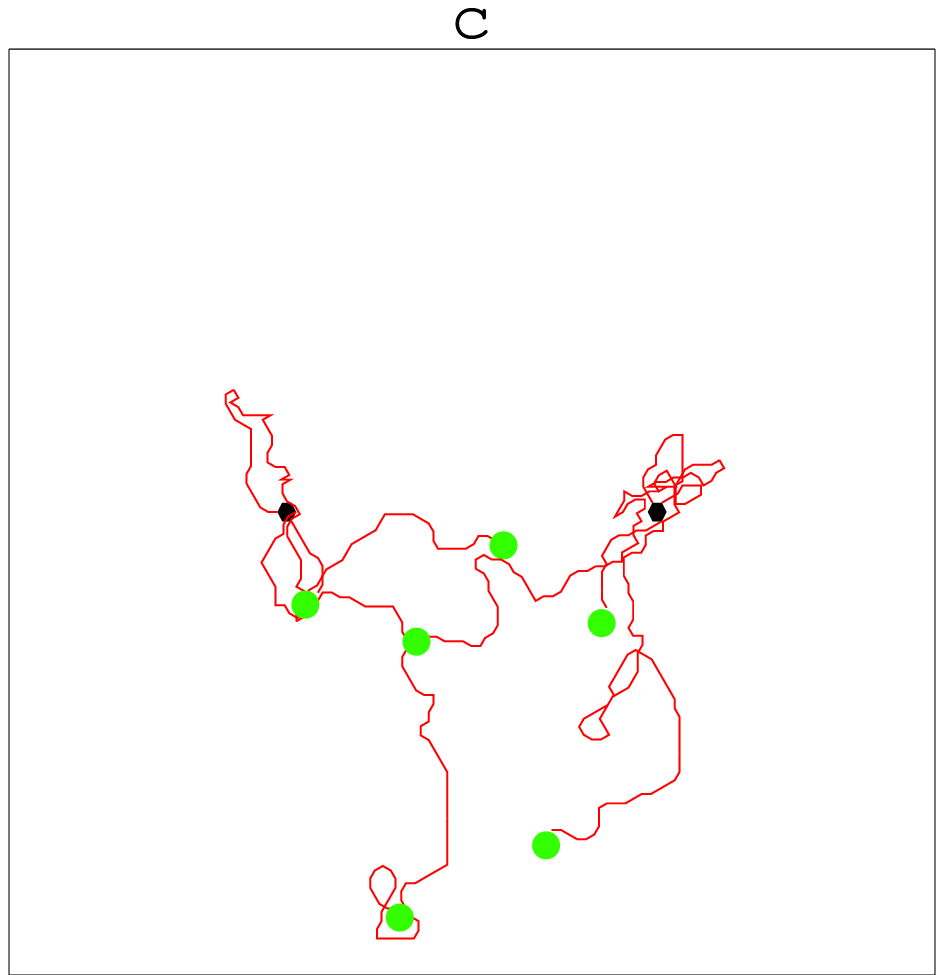,width=8cm,height=8cm,clip=} \psfig{figure=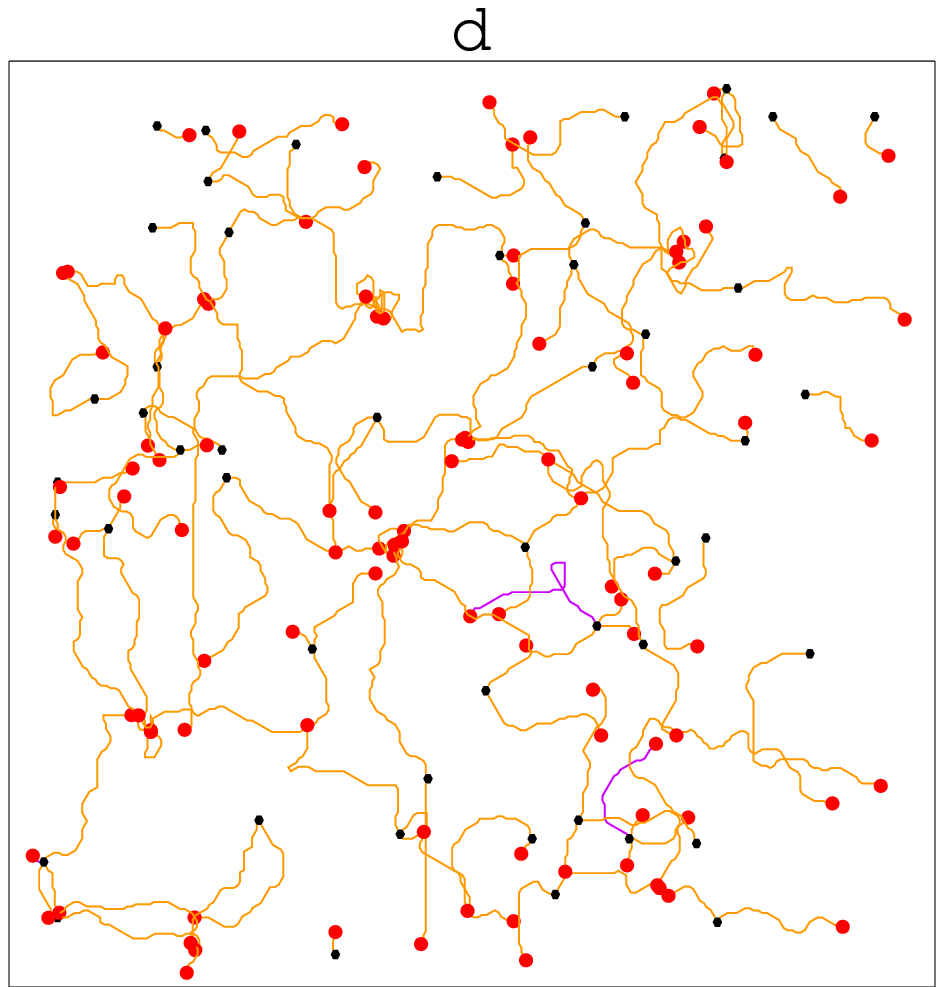,width=8cm,height=8cm,clip=}}
\caption[]{
Simulations of two cells systems :
a. The effect of chemoattractant on the efficiency of navigation. The cell on the right
is a "normal" cell and the cell on the left is a "mutant" cell that does not emit neurites.
Note that even a walker that first migrates away from the target cell navigates towards this cell after
it has been triggered. The contours correspond to different concentrations of the chemoattractant.
b. Self-wiring in a two-cells system. Here both are "normal" cells with identical
partners. The synaptic connections are formed at about half-way between the cells. The wiring
process is very efficient: five out of the six emitted neurites formed connections. The  dots are the "synaptic" connections.
c-d. In figures 2c and 2d we show the effect of 2-fold imposed anisotropy.
(We impose a comb of strips \( a_0 \) wide, \(3a_0 \) for c and $10a_0$ for d, between the strips.) c. Two-cell system.
d. 50-cell system. The growth cones have higher
probability to move along the strips. The resulting pattern agrees with experimental observations \cite{Dwir96}. 
}
\end{figure}

\begin{figure}[t]
\centerline{\psfig{figure=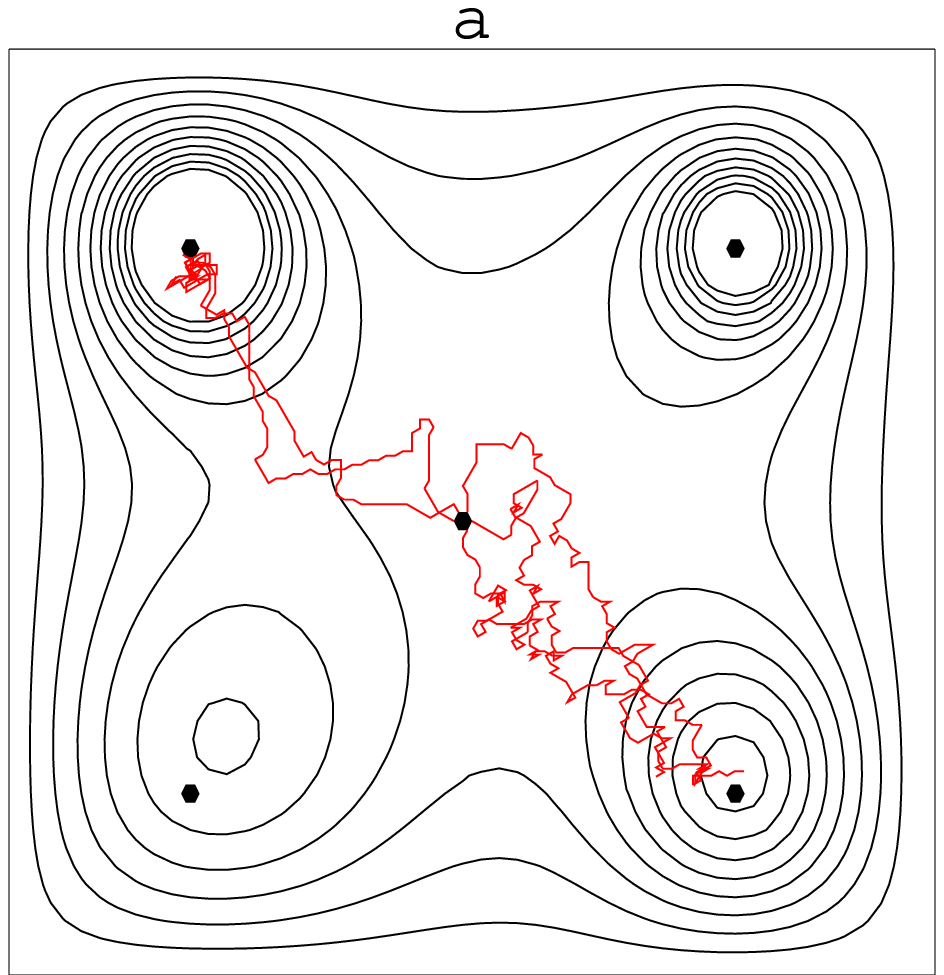,width=8cm,height=8cm,clip=} \psfig{figure=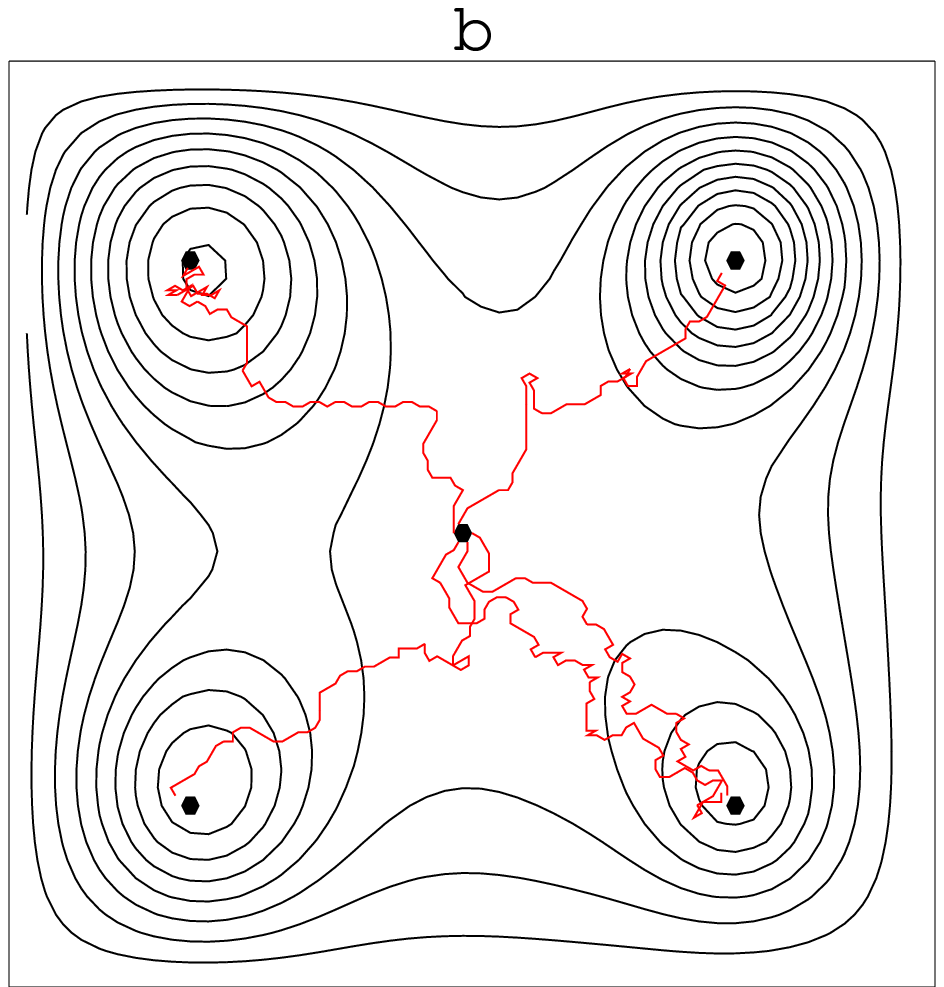,width=8cm,height=8cm,clip=}}
\caption[]{
Simulations of five-cell systems. The central cell is "normal" and the four cells at the
corners are "mutant" cells. The contours are as in Figure 2a.
a. Low rate of "internal" energy "feeding" so that \( l_c \) is much smaller that the
inter-cell distances. In this case the wiring is not efficient as the central cell is wired only
to two of the four neighbors.
b. Higher rate of "feeding" so that \( l_c \) is approximately half the inter-cellular
distance. In this case the wiring is more efficient and the central cell is wired to all its
neighbor cells.}
\end{figure}

\begin{figure}[t]
\centerline{\psfig{figure=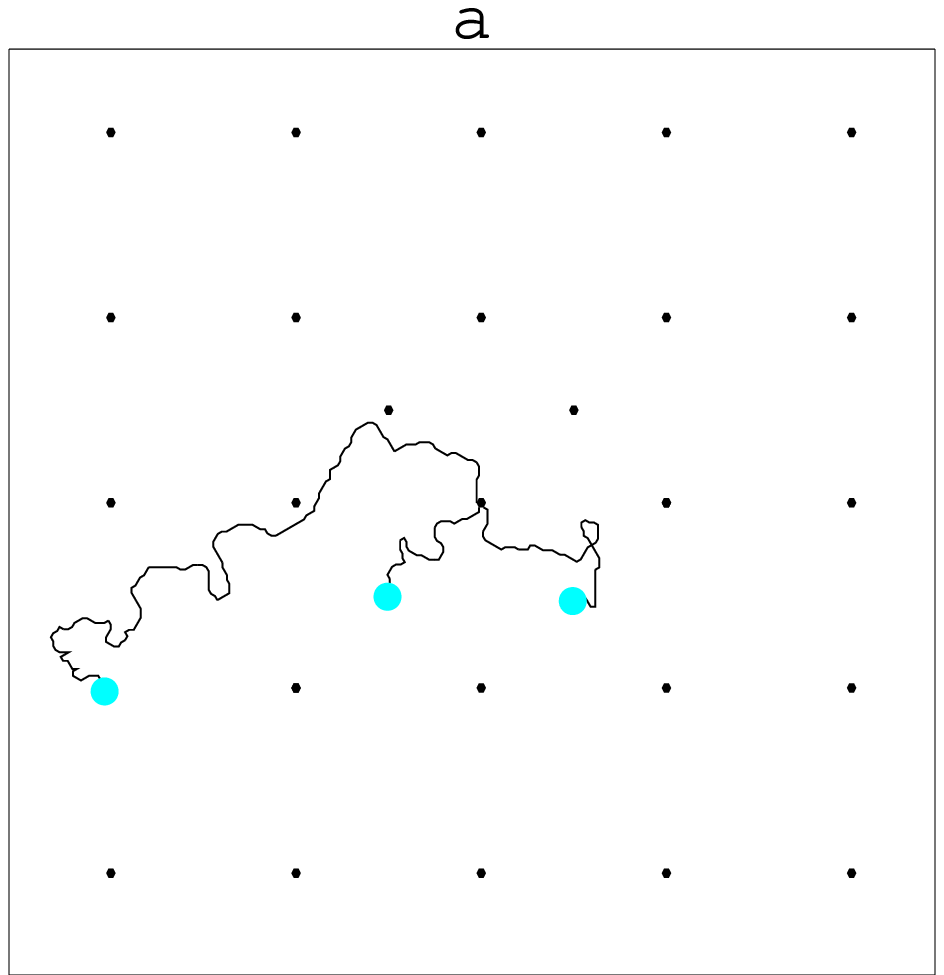,width=8cm,height=8cm,clip=} \psfig{figure=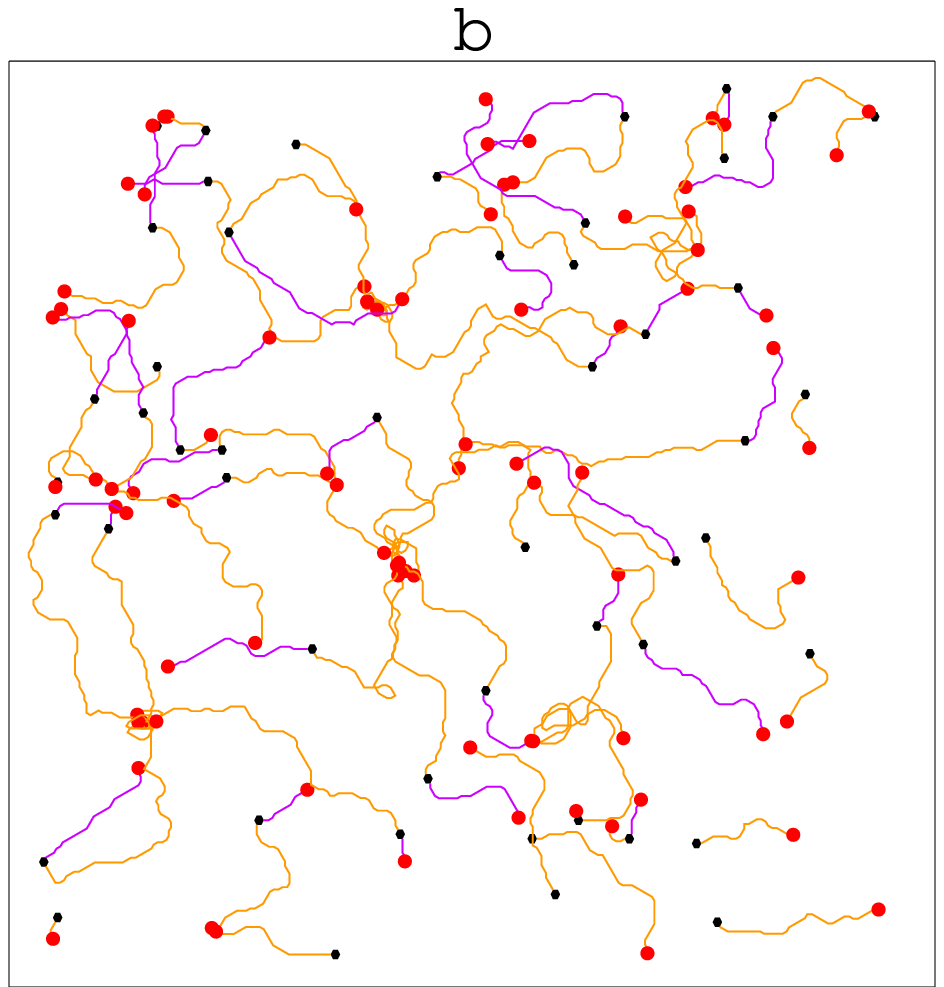,width=8cm,height=8cm,clip=}}
\caption[]{
a. Simulations of a system composed of 30 cells. Only the cell at the center
is "normal" and all other cells are "mutants". The central cell has four nearest neighbor
(NN) cells and eight next nearest neighbor (NNN) cells. 
At the beginning of the growth \( l_c \) is about half the inter-cellular distance. Thus the central cell is wired
only to its NN cells. After The central cell forms two connections the "feeding" rate doubles (doubling of
\(l_c\)). The new neurites navigate to the NNN cells. It demonstrates the manner in
which the soma cell can regulates self-wiring. 
b. Simulation of 100 cell system on a grid of size $200\times200$.
}
\end{figure}

\end{document}